\documentclass[]{spie}  

\usepackage{amsmath,amsfonts,amssymb}
\usepackage{graphicx}
\usepackage[colorlinks=true, allcolors=blue]{hyperref}

\title{Characterization of solar background signatures in IXPE gas pixel detectors through laboratory measurements}

\author[a,b,c]{Simone Pagliarella}
\author[a]{Riccardo Ferrazzoli}
\author[a]{Paolo Soffitta}
\author[a]{Enrico Costa}
\author[a]{Alessandro Di Marco}
\author[a]{Nicolas De Angelis}
\author[a]{Alessandro Lacerenza}
\author[a]{Fabio Muleri}
\author[d]{Stefano Silvestri}
\affil[a]{INAF Istituto di Astrofisica e Planetologia Spaziali, Via del Fosso del Cavaliere 100, 00133 Roma, Italy}
\affil[b]{Tor Vergata University of Rome, Via Della Ricerca Scientifica 1, 00133 Roma, Italy}
\affil[c]{Dipartimento di Fisica, Universit\`a degli Studi di Roma `La Sapienza', P.le Aldo Moro 2, 00133 Roma, Italy}
\affil[d]{Istituto Nazionale di Fisica Nucleare, Sezione di Pisa, Largo B. Pontecorvo 3, I-56127 Pisa, Italy}
 
\authorinfo{Further author information: send correspondence to S.P.: simone.pagliarella@uniroma1.it}

\begin{document} 
\maketitle

\begin{abstract}
We present a laboratory to in-flight comparison of the IXPE background signatures, using the spare Detector Unit (DU), identical to the three flight units, exposed to a $^{60}$Co radioactive source, historically adopted as a proxy for solar-induced X-ray background. $^{60}$Co emits 1.17 MeV and 1.33 MeV photons which undergo Compton scattering in the DU structures, producing a degraded spectrum extending down to the keV range after ADC counts to keV conversion via Fe-55 (charging measurement) calibration. The source was placed on the top and side of the DU; event tracks were reconstructed to extract spectral and morphological observables (track size, energy fraction in the track’s main cluster, number of pixel triggered per event). Laboratory spectra and parameter distributions are compared with IXPE in-flight in-eclipse (Earth occultating the sun) and in-sun backgrounds from two observations (NGC~2110, RCW~86). We discuss how the in-flight and on-ground spectral and modulation properties of the background compare to each other. 
\end{abstract}


\section{INTRODUCTION}
\label{sec:intro}  

The Imaging X-ray Polarimetry Explorer (IXPE, \citenum{2022Weisskopf, Soffitta_2021}) carries three identical Gas Pixel Detectors (GPDs, \citenum{2001Costa,2006Bellazzini}), which combine sensitivity to the polarization of incoming X-rays with imaging and spectroscopic capabilities, reconstructing the photoelectron track produced by each absorbed photon to recover its energy, absorption point, and emission angle. Like any X-ray instrument operated in space, IXPE detectors are affected by an instrumental and environmental background whose intensity and spectral properties can vary with the orbital position and with the solar activity, since charged particles and X-rays of solar origin contribute to the counts recorded by the detectors in addition to the diffuse cosmic and instrumental components. Characterizing this background, and disentangling it from the source signal, becomes especially important for the analysis of faint and extended sources, where background uncertainties can significantly affect the interpretation of the measured polarization degree (PD) and angle (PA). 

Indeed, previous works have investigated the properties of the IXPE background and its relevance for background selection and rejection [\citenum{2023DiMarco}], as well as its dependence on the particle environment encountered along the orbit [\citenum{2021Xie}]; and furthermore, reconciling ground-based expectations with the background actually observed in flight is the subject of ongoing effort within the collaboration [\citenum{2026Soffitta}]. In this context, laboratory measurements performed with a spare Detector Unit, exposed to radioactive sources chosen to mimic the spectral signature of specific background components, provide a controlled and repeatable way to investigate the detector response independently of the variability associated with orbital and solar conditions, and to build reference templates that can later be compared with the background actually observed in flight.

With this aim, in this work we use a $^{60}$Co radioactive source (whose high-energy photons are progressively degraded via Compton scattering down, with the detector housing, to the IXPE energy band) as a laboratory proxy for the solar-induced X-ray background, historically adopted for this purpose within the collaboration. We compare the resulting spectral and track-morphology properties, measured under different illumination geometries, with the in-flight background measured during two IXPE observations selected for their significant exposure to varying solar conditions: NGC~2110 [\citenum{2025Chakraborty}], a nearby active galactic nucleus, and RCW~86 [\citenum{2026Silvestri}], a Galactic supernova remnant.

\section{DATA \& METHODS}

\subsection{Laboratory measurements}

We performed laboratory measurements in a controlled clean-room environment using radioactive calibration sources in order to characterize the IXPE Gas Pixel Detector (GPD) response and provide reference datasets for comparison with in-flight observations. The experimental setup was based on the standard IXPE GPD calibration configuration [\citenum{2022Muleri}], including the detector assembly (DU-FM1), front-end electronics, and data acquisition system operated under stable environmental conditions.

We carried out a preliminary energy calibration using a $^{55}$Fe source: illuminating the detector from above to ensure uniform coverage of the sensitive area: we used the K$\alpha$ emission line at $\sim$5.9 keV [\citenum{2000Knoll}] as a reference to determine the ADC-to-keV conversion factor, monitoring the temporal evolution of the peak position throughout the measurement, to verify the stability of the detector gain. This calibration defines the energy scale adopted for both the laboratory and the in-flight datasets analyzed in this work.

Then, we acquired the main laboratory measurements with a $^{60}$Co radioactive source, which emits two $\gamma$-ray lines at 1.17 MeV and 1.33 MeV with comparable emission probabilities [\citenum{2000Knoll}]. These high-energy photons interact predominantly via Compton scattering, producing scattered photons and secondary electrons that are progressively degraded through multiple interactions within the detector housing and surrounding materials, populating a continuum of deposited energies extending down to the keV range.

\begin{figure}[h!]
    \centering
    \includegraphics[width=0.6\linewidth]{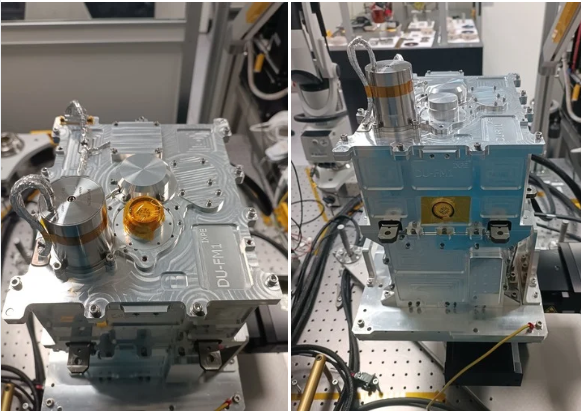}
    \caption{Image of the $^{60}$Co source in the top (left) and side (right) measurements' configuration scheme.}
    \label{fig:configuration_scheme}
\end{figure}

\begin{figure}[h!]
    \centering
    \includegraphics[width=0.35\linewidth]{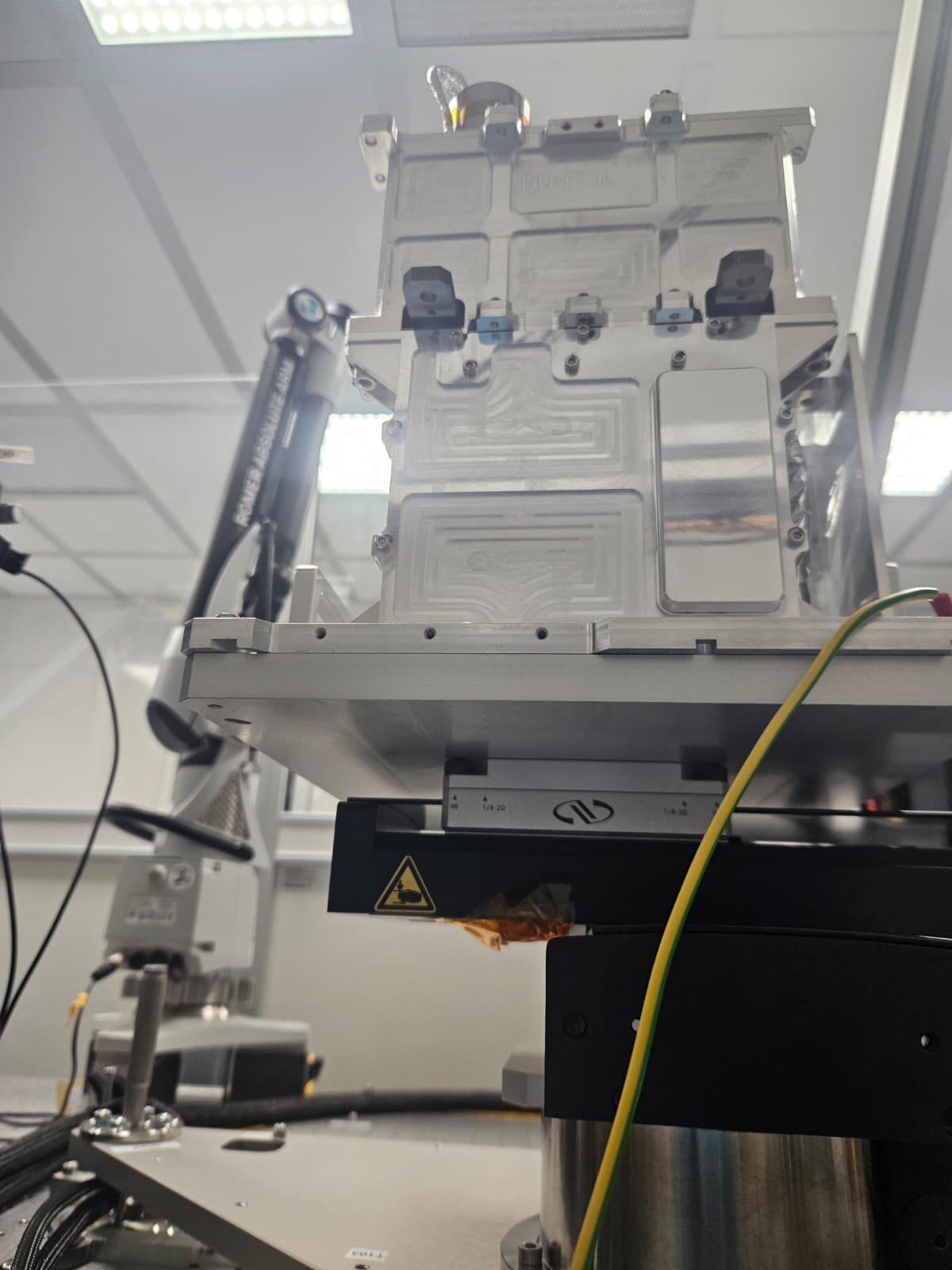}
    \caption{Image of the $^{60}$Co source in the bottom measurement's configuration scheme.}
    \label{fig:configuration_scheme2}
\end{figure}

We acquired the data in three illumination configurations: \textit{top}, \textit{side}, and \textit{bottom} (Figures \ref{fig:configuration_scheme}, \ref{fig:configuration_scheme2}). The \textit{top} and \textit{side} configurations constitute the main reference datasets for the comparison with in-flight observations, while the \textit{bottom} configuration served as a consistency check due to its lower count rate. These configurations allowed us to investigate the dependence of the detector response and reconstructed track morphology on the photon incidence geometry.

\newpage

We processed all laboratory datasets using the standard \texttt{custom gpd} software package, adopted during the IXPE calibration campaign, generating \texttt{quicklook} products that provide a preliminary qualitative overview of the spectral distributions and basic polarimetric properties.\\
We applied a preliminary event selection to reduce spurious events and instrumental artifacts by requiring ADC counts $>$ 1000 and a spatial selection within a 4.2~mm-radius circular region on the detector plane, excluding detector's edges regions affected by non-uniform response [\citenum{2012Soffitta, 2023DiMarco}].\\
We also removed diagnostic events identified through the Level~1 event flags, prior to the analysis.

For each configuration, the analysis focused on three track reconstruction parameters evaluated as functions of reconstructed energy: \texttt{N\_pix} (number of pixels triggered by a certain event), \texttt{Evt\_fra} (ratio between energy in the main cluster and energy of the total track), and \texttt{Trk\_size} (number of pixels above threshold in the main cluster). These parameters describe the morphology and quality of the reconstructed events [\citenum{2021Baldini, 2023DiMarco}] and were used to compare the \textit{top} and \textit{side} illumination configurations.

Finally, we also extracted spectral distributions over the 1-12 keV energy range, with particular emphasis on the nominal IXPE 2-8 keV band. For comparison purposes, we normalized all the spectra to the total number of counts, enabling a direct comparison of the spectral shapes between laboratory and in-flight datasets.

\subsection{In-flight Observations}

We used two IXPE observations of astrophysical sources selected for their significant exposure to varying solar conditions, which can affect the properties of the in-flight background: NGC~2110 (ObsID 03008799), a nearby active galactic nucleus observed between 16-22 and 26-30 October 2024, and RCW~86 (ObsID 02001599, Southwest Rim region), a Galactic supernova remnant observed between 23-30 July 2023, respectively discussed in \citenum{2025Chakraborty} and \citenum{2026Silvestri}. 

We performed the data reduction and analysis using the \texttt{ixpeobssim} software package [\citenum{2022Baldini}]. Additionally, since the track-reconstruction parameters used in this work (N\_pix, Evt\_fra) are only available in the Level~1 event files, we matched events between Level~1 and Level~2 data products using their common TIME column, adding the relevant parameters to the calibrated Level~2 events.

\begin{figure}[h!]
    \centering
    \includegraphics[width=0.6\linewidth]{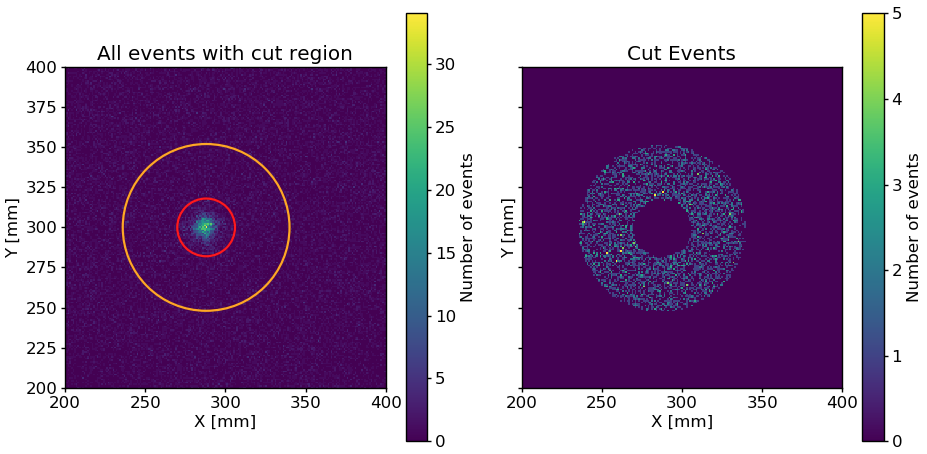}
    \caption{\small Spatial annular region on the detector plane used for background event selection in the NGC~2110 \textit{ineclipse} dataset.}
    \label{fig:cut_NGC2110_ineclipse}
\end{figure}

We selected background events using different strategies for the two observations. For NGC~2110, we defined an annular region surrounding the source on the detector plane, in order to isolate an off-source background sample while minimizing edge effects [\citenum{2023DiMarco}] (see Fig. \ref{fig:cut_NGC2110_ineclipse}). 

For RCW~86, we instead selected time intervals in which the source was fully occulted by the Earth, so that the remaining events are dominated by the background component.\\
In both cases, we further classified the selected events into \textit{ineclipse} and \textit{insun} subsets; corresponding, respectively, to periods in which the spacecraft is shielded from or exposed to direct solar radiation, using the \texttt{xpsun} command of the \texttt{ixpeobssim} pipeline.\\
We, then, applied the same analysis workflow adopted for the laboratory datasets to the in-flight data, including the evaluation of \texttt{N\_pix}, \texttt{Evt\_fra}, and \texttt{Trk\_size} as a function of energy, as well as the extraction of spectral distributions in the 0-12 keV range, with particular emphasis on the nominal IXPE 2-8 keV band.

\section{RESULTS} \label{sec:sections}

\subsection{Laboratory measurements}

The measured count rates for the different illumination configurations were approximately $R_{\mathrm{side}} \sim 1.20$ counts s$^{-1}$, $R_{\mathrm{top}} \sim 0.90$ counts s$^{-1}$, and $R_{\mathrm{bottom}} \sim 0.22$ counts s$^{-1}$. The \textit{top} and \textit{side} configurations were used as the main reference datasets for the comparison with in-flight observations, while the \textit{bottom} configuration served as a consistency check.

\begin{figure}[h!]
    \centering
    \includegraphics[width=1\linewidth]{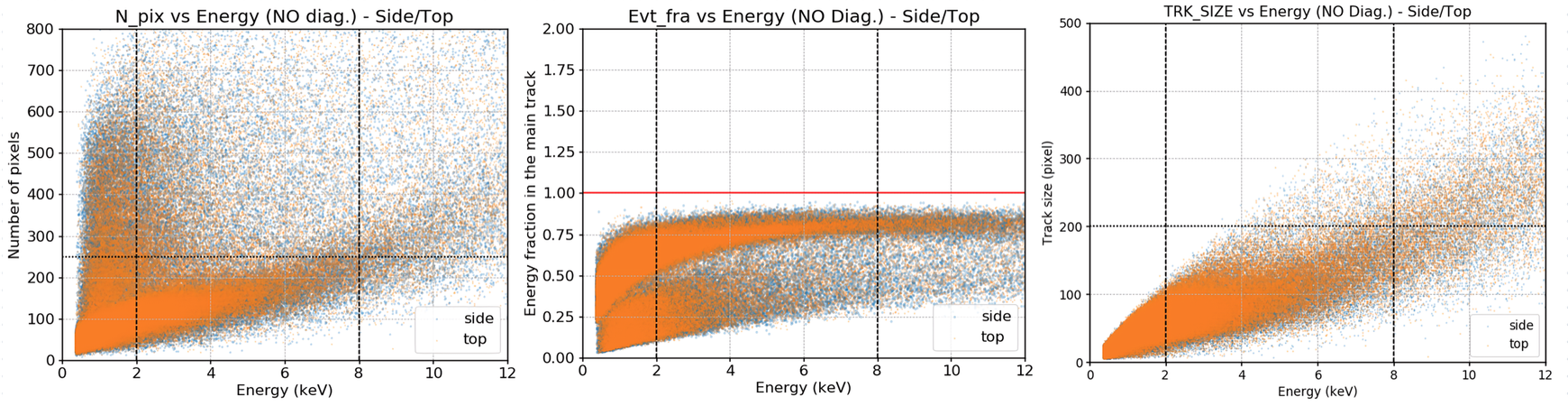}
    \caption{\small Distribution of the \texttt{N\_pix}, \texttt{Evt\_fra}, \texttt{Trk\_size} parameters as a function of energy, comparing side (blue) and top (orange) configurations.}
    \label{fig:param_comparison3}
\end{figure}

Figure~\ref{fig:param_comparison3} shows the distributions of the main track reconstruction parameters, \texttt{N\_pix}, \texttt{Evt\_fra}, and \texttt{Trk\_size}, as a function of reconstructed energy for the \textit{top} and \textit{side} illumination configurations. The three parameters exhibit comparable trends over the analyzed energy range, indicating that the different photon incidence geometries do not significantly affect the reconstructed event morphology.

\begin{figure}[h!]
    \centering
    \includegraphics[width=0.8\linewidth]{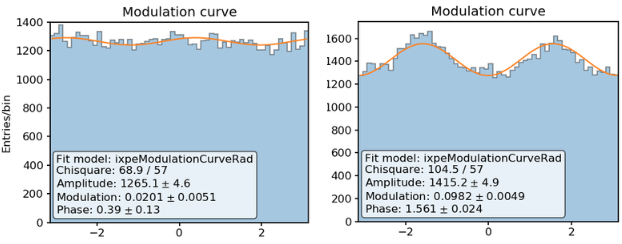}
    \caption{\small Modulation curve for \textit{top} (left panel) and \textit{side} (right panel) configuration setups.}
    \label{fig:modulation_curves}
\end{figure}

A clear difference is instead observed in the modulation response: the \textit{side} configuration exhibits a pronounced modulation due to the preferred photon incidence direction, whereas the \textit{top} configuration is consistent with the expected residual spurious modulation (see Fig. \ref{fig:modulation_curves})[\citenum{2022Rankin}]. This confirms that the modulation behavior is primarily determined by the illumination geometry, while the reconstructed track properties remain largely stable.

\subsection{In-flight comparison}

\begin{figure}[h!]
    \centering
    \includegraphics[width=0.8\linewidth]{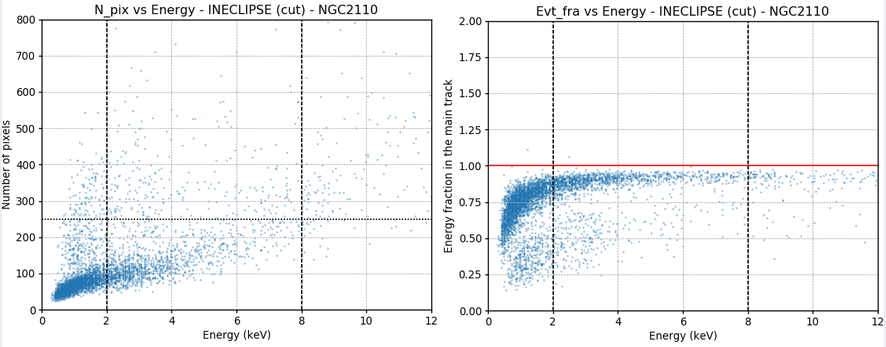}
    \caption{\small Distribution of the \texttt{N\_pix} (left) and \texttt{Evt\_fra} (right) track parameters as a function of energy for the NGC~2110 \textit{ineclipse} background observation.}
    \label{fig:NGC2110_parameters_comparison}
\end{figure}

\begin{figure}[h!]
    \centering
    \includegraphics[width=0.8\linewidth]{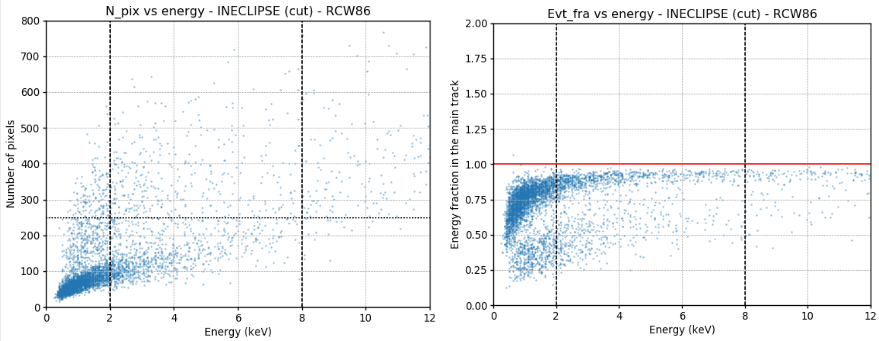}
    \caption{\small Distribution of the \texttt{N\_pix} (left) and \texttt{Evt\_fra} (right) track parameters as a function of energy for the RCW~86 \textit{ineclipse} background observation.}
    \label{fig:RCW86_parameters_comparison}
\end{figure}

The distributions of the main track reconstruction parameters for the NGC~2110 and RCW~86 background datasets (\textit{ineclipse} one) are shown in Figures~\ref{fig:NGC2110_parameters_comparison} and~\ref{fig:RCW86_parameters_comparison}, respectively. For both observations, the \texttt{N\_pix} and \texttt{Evt\_fra} distributions exhibit similar behavior between them, and their trends are also consistent with those obtained from the laboratory measurements. This indicates that the reconstructed track morphology is preserved between the different laboratory setups and both in-flight background datasets.

\begin{figure}[h!]
    \centering
    \includegraphics[width=0.8\linewidth]{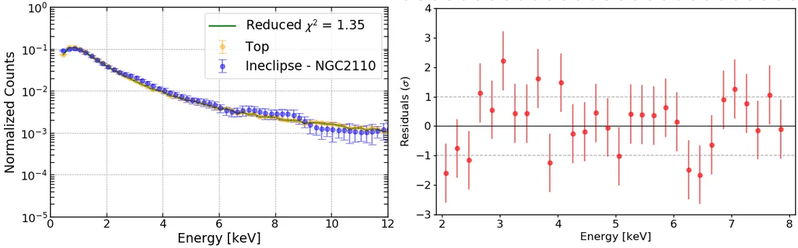}
    \caption{\small Spectral comparison between the laboratory \textit{top} measurement and the NGC~2110 \textit{ineclipse} background dataset, with relative residuals from 2 to 8 keV.}
    \label{fig:NGC2110_spectra_comparison}
\end{figure}

\begin{figure}[h!]
    \centering
    \includegraphics[width=0.8\linewidth]{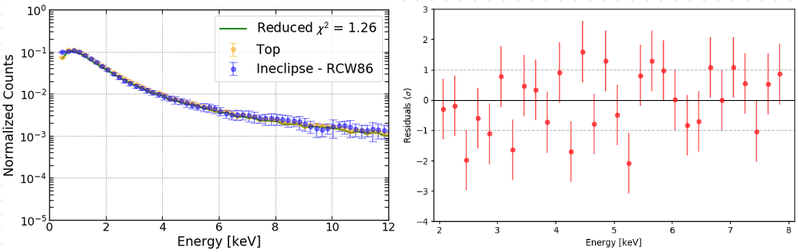}
    \caption{\small Spectral comparison between the laboratory \textit{top} measurement and the RCW~86 \textit{ineclipse} background dataset, with relative residuals from 2 to 8 keV.}
    \label{fig:RCW86_spectra_comparison}
\end{figure}

Beyond this morphological comparison, the main result of the analysis is provided by the spectral comparison between the laboratory measurements and the in-flight background spectra. Figures~\ref{fig:NGC2110_spectra_comparison} and~\ref{fig:RCW86_spectra_comparison} show the comparisons for NGC~2110 and RCW~86, respectively, using the laboratory \textit{top} spectra as reference templates.\\
In all cases, the laboratory spectra reproduce the overall shape of the in-flight background after normalization to the total number of counts.

For NGC~2110, the \textit{top} illumination configuration provides a better agreement with the in-flight background spectrum than the \textit{side} configuration, with reduced $\chi^2$ values of 1.35 and 1.02 in the 1-12 keV and 2-8 keV bands, respectively. The corresponding values for the \textit{side} configuration are 1.86 and 1.53.

We observed a similar behavior for RCW~86, where the \textit{top} configuration yields reduced $\chi^2$ values of 1.26 and 1.10 in the 1-12 keV and 2-8 keV bands, respectively, compared with 2.04 and 2.06 for the \textit{side} configuration. These results indicate that the \textit{top} laboratory configuration provides the closest representation of the in-flight background spectrum for both independent IXPE observations.

\newpage

\section{CONCLUSIONS}

The comparison between laboratory measurements and IXPE in-flight data shows that controlled GPD measurements provide a representative description of the main characteristics of the orbital in-flight background during Sun-influenced observations. The consistency observed in the \texttt{N\_pix}, \texttt{Evt\_fra}, and \texttt{Trk\_size} distributions indicates that the reconstructed event morphology is largely preserved between laboratory and in-flight datasets, despite the different acquisition conditions and photon incidence geometries. At the same time, the different modulation responses observed for the \textit{top} and \textit{side} configurations are consistent with the expected effect of the illumination geometry, confirming that the modulation behavior is primarily determined by the measurement setup.\\
Despite the different acquisition environments and the different background-extraction strategies adopted for the two observations (spatial for NGC~2110, temporal for RCW~86) both the reconstructed event properties and the spectral distributions exhibit a remarkable level of consistency between laboratory and orbital data.

Indeed, the spectral comparison represents the most stringent test of the laboratory measurements as background templates. For both NGC~2110 and RCW~86, the \textit{top} illumination configuration provides the closest agreement with the in-flight background spectra, particularly in the nominal IXPE 2-8 keV energy band. The best agreement is obtained for the \textit{ineclipse} datasets, corresponding to intervals in which the spacecraft is shielded from direct solar radiation by the Earth; a complete characterization of the background observed during \textit{insun} intervals will require further dedicated investigation and additional measurement campaigns.\\
These preliminary results suggest that laboratory measurements can provide a useful reference for the characterization of the IXPE instrumental background [\citenum{2023DiMarco}]. Future analyses, including additional IXPE observations and laboratory campaigns with different radioactive sources, will be required to investigate the background behavior under a broader range of orbital and environmental conditions and complement ongoing efforts to reconcile ground-based expectations with the observed background [\citenum{2026Soffitta}].

\bibliography{sample701} 
\bibliographystyle{spiebib} 

\end{document}